\documentclass[nofootinbib,amsfonts,prd,aps]{revtex4}
\usepackage{url}
\begin{document}

\title{No fermion doubling in quantum geometry}

\author{Rodolfo Gambini$^{1}$, 
Jorge Pullin$^{2}$}
\affiliation {
1. Instituto de F\'{\i}sica, Facultad de Ciencias, 
Igu\'a 4225, esq. Mataojo, 11400 Montevideo, Uruguay. \\
2. Department of Physics and Astronomy, Louisiana State University,
Baton Rouge, LA 70803-4001}

\begin{abstract}
  In loop quantum gravity the discrete nature of quantum geometry acts
  as a natural regulator for matter theories. Studies of quantum field
  theory in quantum space-times in spherical symmetry in the canonical
  approach have shown that the main effect of the
  quantum geometry is to discretize the equations of matter
  fields. This raises the possibility that in the case of fermion
  fields one could confront the usual fermion doubling problem that arises in
  lattice gauge theories. We suggest, again based on recent results on
  spherical symmetry, that since the background space-times will
  generically involve superpositions of states associated with
  different discretizations the phenomenon may not arise. This opens
  a possibility of incorporating chiral fermions in the framework of
  loop quantum gravity.
\end{abstract}
\date{\today}
\maketitle

Loop quantum gravity has provided a non-trivial, anomaly free, finite
theory of quantum general relativity coupled to matter \cite{qsd}. In
it, the quantum geometry plays the role of regulator of matter
fields. Although this is suggested in the general theory, difficulties
in making progress in general have hampered exhibiting detailed
examples.  On the other hand, detailed calculations are possible in
the context of spherically symmetric space-times, where the theory can
be solved completely in closed form \cite{sphericalprl}. In particular
Hawking radiation has been recovered using quantum field theory in
quantum space-time techniques \cite{hawking}. Certain very high
frequency transplanckian modes of the field are suppressed naturally
as a consequence of the discreteness of the quantum space-time. This
opens new possibilities for issues like the backreaction of Hawking
radiation on black holes. What has been seen in the spherically
symmetric case is that if one considers quantum field theories living
on the quantum space-time, one ends up with a Fock-like quantization
where the main ingredient is that the equations of the field theory
become discretized. Although one can consider quantum geometries with
sub-Planckian separations for the vertices of the spin networks (which
play the role of points of the lattice), and therefore have an
excellent approximation to the continuum theory, some important
differences arise. To begin with, one obtains dispersion relations
similar to those of lattices, which suppress the propagation of
certain trans-Planckian modes of wavelength smaller than the lattice
spacing. In turn, this helps eliminate the divergences that arise in
ordinary field theories, for instance, when one computes the
expectation value of the stress energy tensor.

A concern that may arise in this context is what happens when one
considers fermions, particularly chiral ones. As is well known,
fermions on the lattice have the problem of fermion doubling
\cite{susskind}. In fact, fermion doubling has already been noted in
certain models of loop quantum gravity \cite{smolin}. Under very general assumptions, the Nielsen--Ninomiya
no-go theorem makes their appearance inevitable. Let us briefly recall
how this problem arises. The dispersion relation for a fermion on a
one dimensional lattice (or on a spherical quantum space-time) is given by,
\begin{equation}
  \omega_n = \pm \frac{1}{\Delta} \sin\left(l_n\right),
\end{equation}
with $l_n= k_n \Delta$, $-\pi \leq l_n\leq \pi$ and 
\begin{equation}
  k_n=\frac{2 \pi n}{N \Delta} 
\end{equation}
with $N$ the number of vertices in the spin networks of the background
quantum space-time and  $\Delta$ the separation of the vertices of the
spin network. The quantity $-N/2\le n \le N/2$ is an integer that
characterizes the wave number. For small values of $k_n$ one recovers the linear
dispersion relation of fermions in the continuum. 

For small lattice spacings, the frequencies go as $\Delta^{-1}$. So
one will have finite frequencies will correspond to $l_n\to k_n
\Delta$ or $l_n\pm \pi \to \pm k_n \Delta$.  Waves near $l_n\to 0$ will
correspond to long wavelength modes (compared to the Planck scale)
that will correspond to the modes of the fermions in the
continuum, with the usual linear dispersion relation. But what about
the modes near $l_n\to\pm \pi$? They will also have a linear
dispersion relation. In the case
of a free field theory one can simply choose not to populate those
modes and therefore one would recover at low energies the usual
fermion behavior. But in an interactive theory that may not be
possible. Following Susskind \cite{susskind} we can consider an interaction term
proportional to the charge density of fermions $\psi(m)^\dagger
\psi(m)$ with $m$ denoting the position on the lattice. In momentum space this will lead to an interaction term in
the Hamiltonian of the form, in momentum space,
\begin{equation}
\psi(m)\psi^\dagger(m)=  \int_{-\pi}^\pi \exp\left(i m \left(l-l'\right)\right)
    \psi(l')^\dagger \psi(l) dl dl'.
\end{equation}
Such a term is likely to excite a pair with momentum $\pm k_n$ as with
$\left(\pi/\Delta -k_n\right)$ or 
$\left(-\pi/\Delta +k_n\right)$. Generic couplings will all excite
the low frequency modes near $k_n \to\pm \pi/\Delta$.

However, when one is contemplating a quantum field living on a quantum
space-time, only for a measure zero set of quantum space-times will
one encounter the situation above. That set would be quantum states
with a single spin network leading to a single discretization of the
quantum field. Generically, one would have superpositions with
different separations between the vertices of the spin network
$\Delta$ (and more in general superpositions with non-uniform
spacings, here for simplicity we consider superpositions with uniform
spacings). If one considers that case then it will not be true
anymore that an interaction like the one considered above (and generic
ones as well) will connect $k_n=0$ with $k_n=\pm \pi$. Let us consider
how the above interaction gets modified in this case. We consider a
superposition of spin networks of spacing $\Delta$, 
all of which contain a vertex at the point
$x$. Then we have for the previously introduced interaction term,
\begin{equation}
\psi(x)\psi^\dagger(x)=
\int{d \Delta d\Delta'} \int_{-\pi}^\pi
\exp\left(i{x} \left(\frac{l}{\Delta}-\frac{l'}{\Delta'}\right)\right)
\psi_\Delta(l')^\dagger \psi_{\Delta'}(l) dl dl'.
\end{equation}
And substituting $l\to k \Delta$ and $l' \to -\pi +k' \Delta'$, we
get, 
\begin{equation}
\psi(x)\psi^\dagger(x)=
\int{d \Delta d\Delta'} \int_{-\pi/\Delta}^{\pi/\Delta}
\int_{0}^{2\pi/\Delta'}
\exp\left(-i{x} \frac{\pi}{\Delta'}\right)
\exp\left(i{x} \left(k-k'\right)\right)
\psi_\Delta(-\pi+k'\Delta')^\dagger \psi_{\Delta'}(k\Delta) \Delta \Delta' dk dk'.
\end{equation}
And we see that unlike in the previous case, there appears a phase
factor that when integrated over $\Delta'$ will vanish due to its
oscillatory nature. Therefore the coupling between small $k$ and $\pm
\left(\pi/\Delta-k\right)$ vanishes.

In fact, there is not even the possibility of doubling in this
context. If one computes the probability density
$\psi(x)\psi^\dagger(x)$ for low energies, in the integrand the
momenta of the order of inverse lattice scale will not contribute,
whereas in the usual lattice treatment they do, leading to doubling. 


The argument presented is in a two dimensional example, based on
spherical symmetry, so we need to note that at this point this is only
a suggestion. At the moment we do not have a similar framework in
$3+1$ dimensions that would allow us to extend the argument there at
any level of detail. It should be noted that others have argued that
there is no fermion doubling in $3+1$ dimensions based on the use of
random lattices, for instance, in the spin network context
\cite{spinnet}. The mechanism we are suggesting here is a different
one. A challenge in the use of random lattices is how to make sense of
the continuum limit, which is usually achieved through a sum over
lattices. Here we consider superpositions of background quantum
space-times that provide an effective superposition of lattices, but
one can choose the background quantum space-time to approximate a
semiclassical situation and one does not need to sum over all possible
lattices. Given the amount of interest in if loop quantum gravity can
incorporate chiral fermions \cite{interest}, it is encouraging to have
more than one possible mechanism to address the problem. It may also
lead to insights into the chiral anomaly in $1+1$ dimensions \cite{1p1},
were one could in principle carry out explicit calculations in the
context of quantum geometries.

We wish to thank Lee Smolin for posing to us the question about
fermion doubling during a talk on our approach to quantum field theory
in quantum space-time in spherical symmetry which motivated this
work. We also thank him and Jacob Barnett for sharing their work with
us. This work was supported in part by grant NSF-PHY-1305000, funds of
the Hearne Institute for Theoretical Physics, CCT-LSU and Pedeciba.

\end{document}